\documentstyle[epsf,aps]{revtex}
\begin{document}
\title{The Path Integral Monte Carlo Calculation\\
of Electronic Forces
}
\author{Fenghua Zong and D. M. Ceperley}
\address{
National Center for Supercomputing Applications\\
Department of Physics,
University of Illinois at Urbana-Champaign\\
1110 W. Green Street, Urbana IL 61801
}
\date{\today}
\maketitle
\begin{abstract}

We describe a method to evaluate electronic forces 
by Path Integral Monte Carlo (PIMC). 
Electronic correlations, as well as thermal effects, are included 
naturally in this method.
For fermions, a restricted
approach is used to avoid the ``sign'' problem. 
The PIMC force estimator is local and has a finite variance.
We
applied this method to determine the bond length of H$_2$ and the
chemical reaction barrier of H+H$_2\longrightarrow $H$_2$+H. At low
temperature, good agreement is obtained with ground state
calculations. We studied the proton-proton interaction in an
electron gas as a simple model for hydrogen impurities in metals. 
We calculated the force between the two protons at two electronic
densities corresponding to Na ($r_s=3.93$) and Al
($r_s=2.07$) using a supercell with 38 electrons. 
The result is compared to previous calculations.
We also studied the effect of temperature on the proton-proton
interaction. At very high temperature, our result agrees with the Debye
screening of electrons. As temperature decreases, the Debye theory fails
both because of the strong degeneracy of electrons and most importantly,
the formation of electronic bound states around the protons.
\end{abstract}

\section{Introduction}

Forces are a basic quantity needed in understanding microscopic
systems, $e.g.$ they are basic inputs to molecular
dynamics simulations and to predicting the equilibrium structures.  
For a system of electrons and
nuclei in thermal equilibrium, 
it is a very good approximation to assume that the
electrons follow the motion of the nuclei adiabatically. 
The forces exerted on those nuclei due to the fast moving 
electrons are known as the Born-Oppenheimer (BO) forces.
Ehrenfest\cite{ehrenfest} first related the force to the 
expectation value of the
gradient of the potential which led to the Hellmann-Feynman
theorem\cite{feynman}.
Accurate results have been obtained using this theorem
within the framework of the local density functional theory (LDA)\cite{martin}. 
Unlike the energy, 
the force is directly related to the optimized geometry ($i.e.$
when $\vec F=0$) and it allows one to probe every single nucleus in
the system;
one can determine the forces on each nucleus and optimize their
individual positions concurrently. In conjunction with the total energy,
the force can be used to help to construct the potential energy surface. With
energies known at certain grid points, a more accurate fit
can be obtained if one knows the derivatives at those points also.

Quantum Monte Carlo (QMC) methods are capable of 
treating many-body effects directly, which
is essential in cases where the electron correlations are important. 
The computational demand of QMC
scales as $N^3$\cite{mitas} or less where $N$ is the number of particles, 
while other methods,
which depend on a complete representation of the many-body
wave function or density matrix, such as configuration interaction (CI), 
have an exponential
dependence on the size of the system.
For QMC simulations,
the total energy of the system usually has a variance  
which is proportional to the 
size of the system, making it difficult to distinguish the
contribution of a single particle and the effect of a local
displacement of nuclear positions. 
Consequently, the ability to calculate forces with QMC methods
provides not only a good test of the accuracy of the commonly used LDA
calculations but also an accurate 
many-body approach that could be applied to extended systems.

The calculation of forces with QMC methods is a long standing important
problem.  
There has been some progress\cite{reynolds} in the calculation of forces 
with both
variational Monte Carlo (VMC) and diffusion Monte Carlo (DMC),
however, they are still limited to relatively small systems, for
example, LiH\cite{vrbik}, BH\cite{umrigar} and CuH\cite{belo}.

The most straight forward approach to estimate forces would be to
calculate the total energy difference between two
sets of nuclear configurations which are close together.
Then the force, for example the force component
in $\hat y$ direction on nucleus $i$, can be approximated as:
\begin{equation}
\label{1}
F_{iy}=-\frac{\partial E}{\partial X_{iy}}
\simeq-\frac{E(\vec X+\Delta X_{iy}\hat y)-E(\vec X)}{\Delta X_{iy}}
\end{equation}
where $\vec X$ is the position of all nuclei and
$\vec X_i$ is for nucleus $i$. However, due to the statistical
nature of Monte Carlo, the energy estimation is always
associated with a statistical error $\sigma_E$ and the error for
the force is then $\sigma_F=\sqrt{2}\sigma_E/\Delta X_{iy}$ for 
independent sampling of $\vec X$ and $\vec X+\Delta\vec X$. 
Hence the error diverges as
$\Delta X_{iy}\rightarrow 0$ making this simple approach impractical.

The Hellmann-Feynman theorem expresses the force as the expectation value of
the potential gradient with respect to the wave function $\Psi$:
\begin{equation}
\vec F=-\langle\Psi| \nabla_X V(\vec X)|\Psi\rangle.
\end{equation}
This approach is well suited for Hartree-Fock or LDA type calculations
where the trial wave functions are the eigenfunctions of the Hamiltonian
without statistical fluctuations.
However, as illustrated in Ref. \onlinecite{reynolds}, 
the variance of this force estimator is
infinite for Coulomb systems because of the $1/r$ behavior of the
potential as electrons approach a nucleus. Consequently, it is not
possible to get a reliable estimation of the force by Monte Carlo
methods using this estimator.

There are other analytical derivative methods for QMC. For example, one
can explicitly carry out the derivative of the variational or diffusion
Monte Carlo energy estimator
\cite{reynolds}:
\begin{equation}
E=\frac{\int\Phi_0E_L\Psi dR}{\int \Phi_0\Psi dR}
\end{equation}
and express the force as a function of $\nabla_X \Phi_0$, $\nabla_X E_L$,
and $\nabla_X\Psi$. 
Here $\Psi$ is the trial wave function and $R$ is electronic
coordinate.
Note $\Phi_0$, which is the exact ground state wave function of the
system, is unknown. One either needs to find a good approximation to 
$\nabla_X \Phi_0$
or use further Diffusion Monte Carlo walks to calculate it.
Also, it may be difficult to determine $\nabla_X\Psi$. These difficulties
have prevented routine calculations of forces with QMC.

In this paper, we formulate the force as the 
derivative of the Born-Oppenheimer free energy with respect to
the nuclear coordinates, and then
evaluate the derivative with a Path Integral Monte Carlo (PIMC)
technique. The force estimator is local and easy to compute. It also has
a much smaller variance than that of Eq. (\ref{1}). In the following, we 
first review the basic formulation of PIMC and then show how the force
is computed.
We apply this method to the molecules H$_2$ and H$_3$, and
find good agreement between PIMC results at low temperature with those
of accurate ground state calculations. To demonstrate the method works
in an extended system, we study the
proton-proton interaction in an electron gas, and compare it to
LDA calculations.

\section{Path Integral and Force Calculation}

Path Integral Monte Carlo is a powerful computational technique which is
capable of simulating boson systems\cite{cep95} exactly and fermions
\cite{cep96} 
accurately. Besides total energy, many properties of the system, such as
pair correlation function, specific heat, pressure, momentum
distribution, and the boson super fluid density have been calculated. In
this paper, we show how one can compute the electronic forces with
PIMC as well.

Consider a system of $N$ non-relativistic particles (electrons and
nuclei) interacting via
Coulomb potential. The Hamiltonian is:
\begin{equation}
{\cal H}=-\sum_{i=1}^N\lambda_i\nabla_i^2+\sum_{i<j}\frac{e_ie_j}{r_{ij}}
\end{equation}
where $\lambda_i=\hbar^2/2m_i$. We will use atomic units throughout the
paper: the unit for length is bohr and the unit for energy is
hartree. In these units, the electron charge $e_i=-1$ and the
inverse electron mass $\lambda_i=1/2$.

In principle, all the properties of an equilibrium system at a finite
temperature $T$ can be
determined from the thermal density matrix, which, 
for a Boltzmann system, 
can be expressed in the position representation as: 
\begin{equation}
\rho(R,R';\beta)=\langle R|e^{-\beta{\cal H}}|R'\rangle
\equiv \left\{\prod_i(4\pi\lambda_i\beta)^{-3/2}\exp\left [-\frac{(r_i-r_i')^2}
{4\lambda_i\beta}\right ]\right \} \exp[-U(R,R';\beta)]
\end{equation}
where $R=\{\vec r_1,\ldots,\vec r_N\}$ 
is the set of all particle coordinates, $\beta=1/k_BT$ is the
inverse of the temperature. The
expectation value of an operator ${\cal O}$ is:
\begin{equation}
\langle{\cal O}\rangle
=\frac{1}{Z}\int dRdR'\rho(R,R';\beta)\langle R|{\cal O}|R'\rangle
\end{equation}
where $Z=\exp(-\beta {\cal F})=\int dR\rho(R,R;\beta)$ is the partition function
and ${\cal F}$ is the free energy.

The calculation of many-body density matrix at a finite temperature is
done by expanding it in terms of density
matrices at higher temperatures:
\begin{equation}
\rho(R^0,R^M;\beta)=\int\cdots\int dR^1dR^2\cdots dR^{M-1}
\rho(R^0,R^1;\tau)\cdots\rho(R^{M-1},R^M;\tau)
\end{equation}
where $M$ is the number of time 
slices and $\tau=\beta/M$ is called the time step.

It is much easier to obtain a good approximation to the high temperature
density matrix since the system behaves like a classical system at
high temperature. The pair product approximation\cite{cep95} 
has been shown
to give much smaller errors in the density matrices compared with 
the primitive approximation, so a much larger time step
can be used. In this approach, one solves the exact action
$u_2(r_{ij}, r_{ij}';\tau)$ for a pair of particles and uses:
\begin{equation}
U(R, R'; \tau)=\sum_{i<j}u_2(r_{ij}, r_{ij}';\tau).
\end{equation}
Errors occur only when three or more particles come close to each other.
In our simulation, we used a matrix squaring method\cite{cep95} to
numerically calculate the high temperature pair action.

Because of their relevance, here we review a few properties of the
Coulomb pair action\cite{pollock}. The classical limit of the action is
$\tau v(r)$ where $v(r)$ is the Coulomb potential between the two
particles.
For large $r_{ij}$ or high temperature, one can expand the action in
powers of $\hbar$ or $\tau$ (Wigner-Kirkwood expansion) to get on the
diagonal:
\begin{equation}\label{3'}
u_2(r_{ij},r_{ij};\tau)=\frac{\tau e_ie_j}{r_{ij}}
-\frac{\tau^3(e_ie_j)^2(\lambda_i+\lambda_j)}{12r_{ij}^4}
+{\cal O}(\tau^4).
\end{equation}
At small $r_{ij}$, quantum effects smooth
the divergence of the 
Coulomb potential. As a result, the pair action
and its coordinate derivative are finite. 
The cusp condition
of the Bloch equation gives the slope at the origin:
\begin{equation}\label{3}
\lim_{r_{ij}\rightarrow 0}\frac{\partial u_2(r_{ij},r_{ij}';\tau)}
{\partial r_{ij}}
=-\frac{e_ie_j}{2(\lambda_i+\lambda_j)}.
\end{equation}

Now let us take into account fermi statistics.
Without magnetic fields, the Hamiltonian is
independent of spin and
$s_z$ is a good quantum number. One can treat
identical particles with different $s_z$ as different species and apply
quantum statistics only to electrons of the same spin.
Let ${\cal P}$ be the permutation operator of particle labels.
Then for each spin state\cite{cep96}:
\begin{equation}
\rho_{F}(R,R';\beta)=\frac{1}{N!}
\sum_{{\cal P}}(-1)^{\cal P}
\rho({\cal P} R,R';\beta)
\end{equation}
where $\rho_F$ is the fermion density matrix
and $N$ is the number of electrons in that spin state.

Because odd permutations of fermions 
contribute a minus sign, a direct summation as for bosons
will result in an exceedingly low efficiency as $\beta$ or $N$ 
increase\cite{cep96}.
The Restricted Path Integral Monte Carlo
(RPIMC)\cite{cep96} solves this problem by only allowing paths which do
not cross the nodal surface of the fermion density matrix. 
The nodes are determined by $\rho_F(R^t,R^*;t)=0$.
$R^*$, the reference point, is a special point on the path.
It is the value of the density
matrix with respect to the reference point that restricts the paths.
If one knows the
exact $3N-1$ dimensional nodes, the RPIMC method is exact. In practice, 
the nodes are not known, one
introduces a trial density matrix and uses its nodes instead. In this
paper, we use
the nodal surface of non-interacting particle
systems. This has been shown
to give accurate simulations of hydrogen
plasma\cite{pierleoni} and liquid $^3$He\cite{cep92}.
The success of such a seemingly simple restriction can be
understood if one takes a more realistic pair product density matrix and
applies the end-point approximation, one finds the nodal surface is
exactly that of the non-interacting system\cite{cep96}.
The off-diagonal corrections to the nodes scale as ${\cal O}(t^2)$, they
are important only at fairly low temperatures because the leading
kinetic energy contribution is ${\cal O}(t^{-1})$.

RPIMC gives an additional contribution to the action 
due to the restriction on the
crossing of the nodal surface\cite{cep96}. 
Locally the nodes can be approximated as
hyper-planes. For small time step $\tau$, 
the nodal action is:
\begin{equation}
u_N(R^0,R^M;\beta)=-\sum_{i=0}^{M-1}
\ln\left [1-\exp\left (-\frac{d^id^{i+1}}{\lambda\tau}\right )\right ]
\end{equation}
where $d^i$ is the distance of $R^i$ to the nearest nodes.

Now return to the problem of computing the quantum forces:
consider a system of $N_n$ nuclei and $N_e$ electrons. 
We fix the position of nuclei (no nuclear kinetic energy) and calculate the
Born-Oppenheimer
free energy of the electrons as a function of the nuclear coordinates:
\begin{equation}
{\cal F}\equiv {\cal F}(\{\vec X_1,\cdots,\vec X_{N_n}\};\beta).
\end{equation}
The force exerted on the nuclei in thermal equilibrium is:
\begin{equation}
\vec F=-\nabla_X{\cal F}=\frac{1}{\beta Z}\nabla_XZ
\end{equation}
where $Z$ is the partition function and 
$\nabla_X$ only acts on nuclear coordinates.

Differentiating the path integral expression for the partition
function:
\begin{eqnarray}
\vec F=&&-\frac{1}{\beta Z}\int\cdots\int 
dR^0\cdots R^{M-1}\left [\rho(R^0,R^1;\tau)\cdots
\rho(R^{M-1},R^M;\tau)\times\sum_{i=0}^{M-1}
\frac{\partial U(R^i,R^{i+1};\tau)}{\partial \vec X}\right ]
\\
&&\equiv -\frac{1}{M\tau}\left \langle\sum_{i=0}^{M-1}
\frac{\partial U(R^i,R^{i+1};\tau)}{\partial \vec X}\right \rangle 
\end{eqnarray}
where $\langle\ldots\rangle$ denotes an average over the paths, $R^0= R^M$
and $R^i$ are electronic coordinates only. The nuclear coordinates are
not explicitly written in the action because they are independent of the
time slice and fixed during the simulation.

We arrive at the formula for the force estimator by expanding the action
in terms of sums over pairs of charged particles:
\begin{equation}\label{4}
\vec F=-\frac{1}{M\tau}\left \langle\sum_{i=0}^{M-1} \sum_{kl}
\frac{\partial u_2(r^i_{kl},r^{i+1}_{kl};\tau)}
{\partial \vec X}\right \rangle.
\end{equation}
Note here $k$ and $l$ run over nuclear indices also.
For small $r$ or $r'$, both $u_2(r,r';\tau)$ and $\partial
u_2/\partial r$ (Eq. \ref{3}) have finite values. At large $r$, the action
approaches $\tau e_ie_j/r$. There is no divergence in this force
estimator and thus the error 
is finite in contrast to estimators based on the Hellman-Feynman
theorem.

Fermion nodes cause an additional contribution to the force. 
The electronic nodal surface can
depend on the positions of nuclei, and so does the distance of
$R^i$ to the nodes:
\begin{equation}
d^i=d^i(\{\vec X_1,\cdots,\vec X_{N_n}\}).
\end{equation}
When taking the derivative of
the action with respect to nuclear coordinates, the change of the nodal
surface will contribute a force on the nuclei: 
\begin{eqnarray}
\vec F_N=&&-\frac{1}{\tau}\left \langle\frac{\partial u_N(R^i,R^{i+1};\tau)}
{\partial \vec X}\right \rangle  \\
=&&-\frac{1}{\lambda \tau^2}\left \langle\left [\exp\left (
\frac{d^id^{i+1}}{\lambda\tau}\right )-1\right ]^{-1}
\frac{\partial(d^id^{i+1})}{\partial \vec X}\right \rangle.
\end{eqnarray}
However, this term vanishes with
non-interacting nodal surfaces because the nodal
positions are independent of the nuclear coordinates.

We now investigate how the computer time will depend on the time step.
In general, the variance of the force estimator, $\sigma^2_F$,
is a function of the time step $\tau$. Because
of the high temperature approximations introduced in a PIMC simulation, 
the force is only exact in the limit of $\tau\rightarrow 0$. 
Hence the behavior of the variance at small $\tau$ affects the
overall efficiency of a calculation. To understand the dependence of
$\sigma^2_F$ on $\tau$, we consider a typical, yet simple system: a
hydrogen atom. Assuming independent samples of slices on the path,
the variance of the force is:
\begin{equation}\label{variance}
\sigma^2_F=\frac{1}{\tau^2}\left \langle\left(\frac{\partial u(r^i,r^{i+1};\tau)}
{\partial  \vec X}\right)^2\right \rangle-\frac{1}{\tau^2}
\left(\left \langle\frac{\partial u}{\partial \vec X}\right \rangle\right)^2
\end{equation}
where $r^i$ is the relative coordinate between the electron and the
proton at time step $i$. 
The second term in the above equation is the square of the force
(zero in this case)
and is independent of $\tau$, so we only need to estimate the first
term. 

Both the small $r$ (Eq. \ref{3}) and large $r$ (Eq. \ref{3'}) 
limit for the action is known. We
approximate the derivative of the action as:
\begin{equation}
\frac{\partial u(r,r;\tau)}{\partial r}=\left\{
\begin{array}{ll}
1&|r|\langle r_c\\
\tau/|r|^2&|r|\geq r_c
\end{array}
\right.
\end{equation}
where $r_c=\sqrt{2\lambda\tau}$, the thermal De Broglie wavelength,
is the radius inside of which quantum correction is important.
The variance is thus roughly:
\begin{eqnarray}
\sigma^2_F
\approx 
&&\frac{1}{\tau^2Z}\int dr\rho(r,r;\beta)
\left (\frac{\partial u(r,r;\tau)}{\partial \vec X}\right )^2\\
\propto &&\tau^{-1/2}+{\cal O}(1). \label{variance1}
\end{eqnarray}
The error will diverge when $\tau\rightarrow 0$, but only very slowly, as
$\tau^{-1/4}$.

\begin{figure}
\begin{center}
\leavevmode
\epsfbox{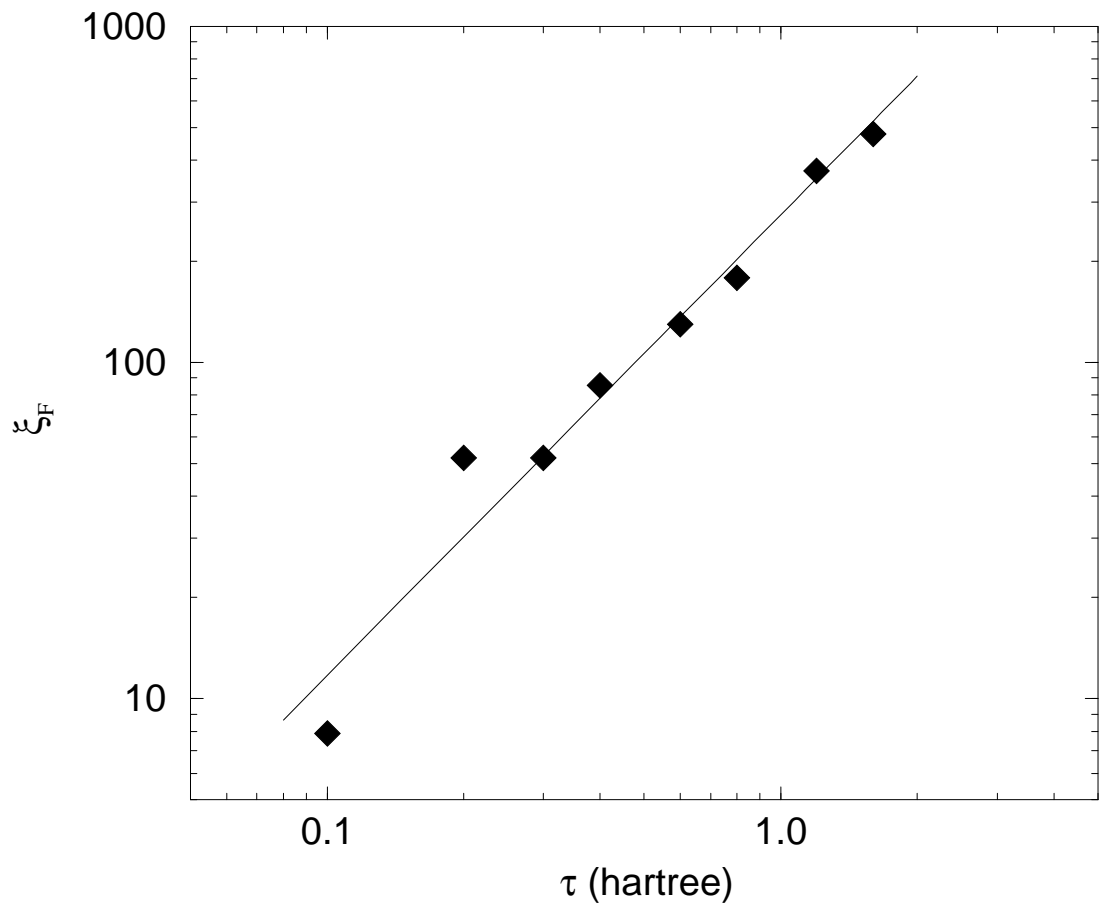}
\end{center}
\caption{The efficiency $\xi_F$ of the PIMC force calculation for a
H$_2$ molecule as a function of time step $\tau$
at an inverse temperature $\beta=19.2$ a.u.
in units of bohr$^2$/hartree$^2$second on a
SGI/CRAY Origin2000 computer. 
The solid line is the power law fit $\tau^{1.4}$.}
\label{H2eff}
\end{figure}

The above estimate does not take into account how quickly independent
samples can be generated or how adjacent time slices are correlated, 
so we performed an empirical study of the
efficiency using our PIMC code.
The efficiency of the PIMC force calculation is defined as:
\begin{equation}\label{eff}
\xi_F=\frac{1}{\sigma_F^2PT}.
\end{equation}
It measures how
quickly the variance of the force, $\sigma^2_F$, decreases 
as a function of computer time.
Here $P$ is the number of Monte Carlo steps and $T$ is the computer
time per step. Fig. (\ref{H2eff}) shows the efficiency as a function of
$\tau$ for the PIMC simulation of a H$_2$ molecule. We found that
$\xi_F$ scales as $\tau^{1.4}$. 

\begin{figure}
\begin{center}
\leavevmode
\epsfbox{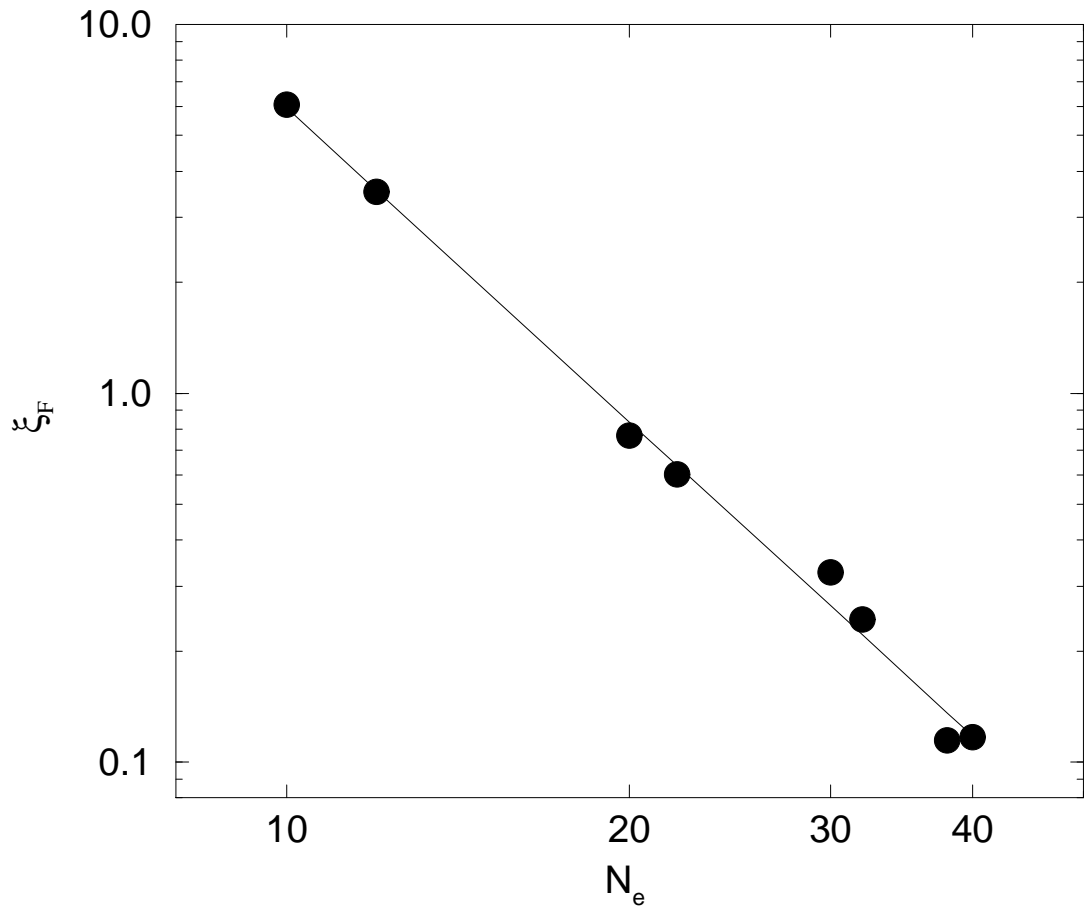}
\end{center}
\caption{The efficiency $\xi_F$ of the PIMC force calculation as a
function of the total number of electrons. The system consists of two
protons and $N_e$ electrons as
discussed in Fig. (\protect\ref{size}).
The solid line is the power law fit $N_e^{-2.8}$.}
\label{Ne_eff}
\end{figure}

The only terms that contribute in Eq. (\ref{4}) are terms involving the
nucleus in question and another charged particle. 
The dominant contributions are local. 
The force on a nucleus mostly comes from nearby
electrons, hence the force variance is mainly due to 
nearby electronic paths and independent of the total number of
electrons.
One can preferentially move electrons that are near
to the nucleus and thereby obtain a better overall efficiency of the
force calculation. We performed PIMC simulations of H impurities in an
electron gas (see section IV) and found a power law behavior of the
efficiency as a function of the number of electrons $N_e$. As shown in
Fig. (\ref{Ne_eff}), $\xi_F\sim N_e^{-2.8}$.

To evaluate the path average in Eq. (\ref{4}), 
one samples the path space with multi-level
Metropolis method, the level is chosen so that the diffusion of
paths in both the coordinate and the permutation space is maximized.
This is discussed in detail in Ref.\onlinecite{cep95,cep96}.
Typically a path segment of $4\sim 16$ slices is moved at the same time.
The permutation space is sampled by cyclic exchange of the labels of 
three particles followed by a path move (for RPIMC,
a two particle permutation gives a minus sign and is not allowed). 
This achieves ergodic sampling of the permutation space.

\section{Force Calculation for H$_2$ and H$_3$}

To test the above approach, we apply it to the
H$_2$ and H$_3$ molecules, since very accurate 
results for these systems are known.
PIMC is a finite temperature method and it is well known that an
isolated Coulomb system at a finite temperature would self-ionize.
To circumvent this problem, we place the molecule in a
periodic cube. When an electron ionizes and moves out of the
simulation cell from one side, it will reenter from the other
side and be captured again by the molecule. The properties of the 
ground state and low excited states of the system are unaffected if the
cell is large enough because the wave functions corresponding to these
states are very small at the cell boundary. 
As a result, these states hardly
feel the existence of the periodic boundary condition and the wave functions
are unchanged. For higher excited states and continuous states which are
more spread out in space, wave functions from adjacent cells overlap
with each other. Those states and their energy spectrum are thus changed
in such a way as to prevent the ionization.
We are only concerned with the low temperature properties of
the system, corresponding to the lowest states of the molecule.
The periodic boundary condition is well suited for this purpose.
The minimum image convention\cite{allen} was used in calculating the
Coulomb interaction. 
No long range contributions from images are included.

\begin{figure}
\begin{center}
\leavevmode
\epsfbox{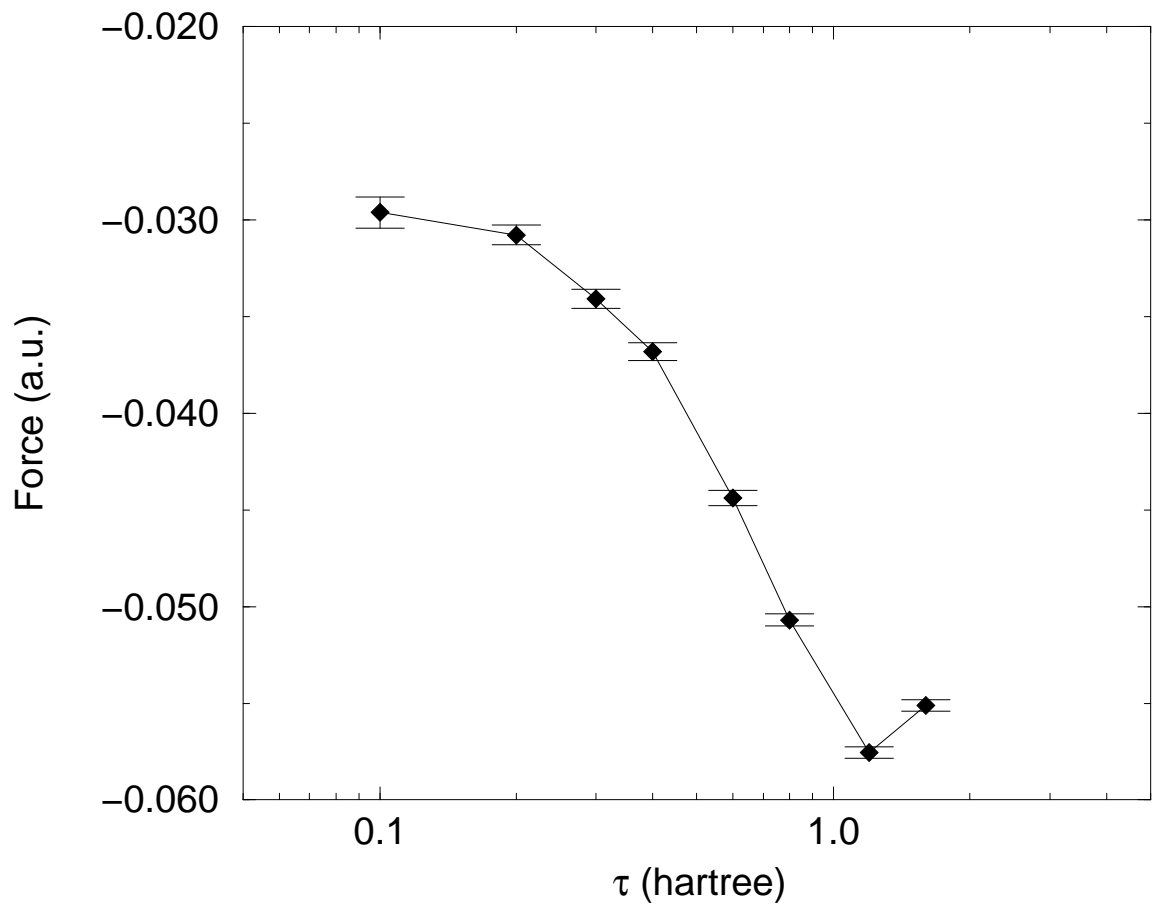}
\end{center}
\caption{The time step dependence of the PIMC force calculation for a
H$_2$ molecule at an inverse temperature of $\beta=19.2$ a.u. The proton
separation is $1.5$ bohrs. The zero temperature result for the force
is $F=-0.031$ a.u.\protect\cite{kolos}}
\label{H2err}
\end{figure}

The ground state of hydrogen molecule (H$_2$) has two electrons with
opposite spin and can be simulated as distinguishable particles. The
first excited electronic state($b ^3\Sigma^+_u$) 
has an energy of 0.39 above the ground state at a proton-proton distance
of $a=1.4$.
At a temperature of $T=0.026$ hartrees ($\beta=38.4$), we reproduce the
H$_2$ ground state energy within statistical accuracy of $0.001$ hartrees.
The time step dependence of the PIMC force 
on $\tau$ is plotted in Fig. (\ref{H2err}). 
At $\tau\leq 0.2$
the PIMC result is close to the zero temperature result of $F=-0.031$ a.u.
within error bars.
In the following, we use a time step of $\tau=0.2$ with
$196$ time slices. 
We choose the simulation cell to be a cubic box with length $L=20.0$. 
Calculations show that
for $L\geq 20$, the boundaries do not affect the electron-proton
pair correlation function at $L/2$ and a
convergent result with respect to cell size is obtained.

\begin{figure}
\begin{center}
\leavevmode
\epsfbox{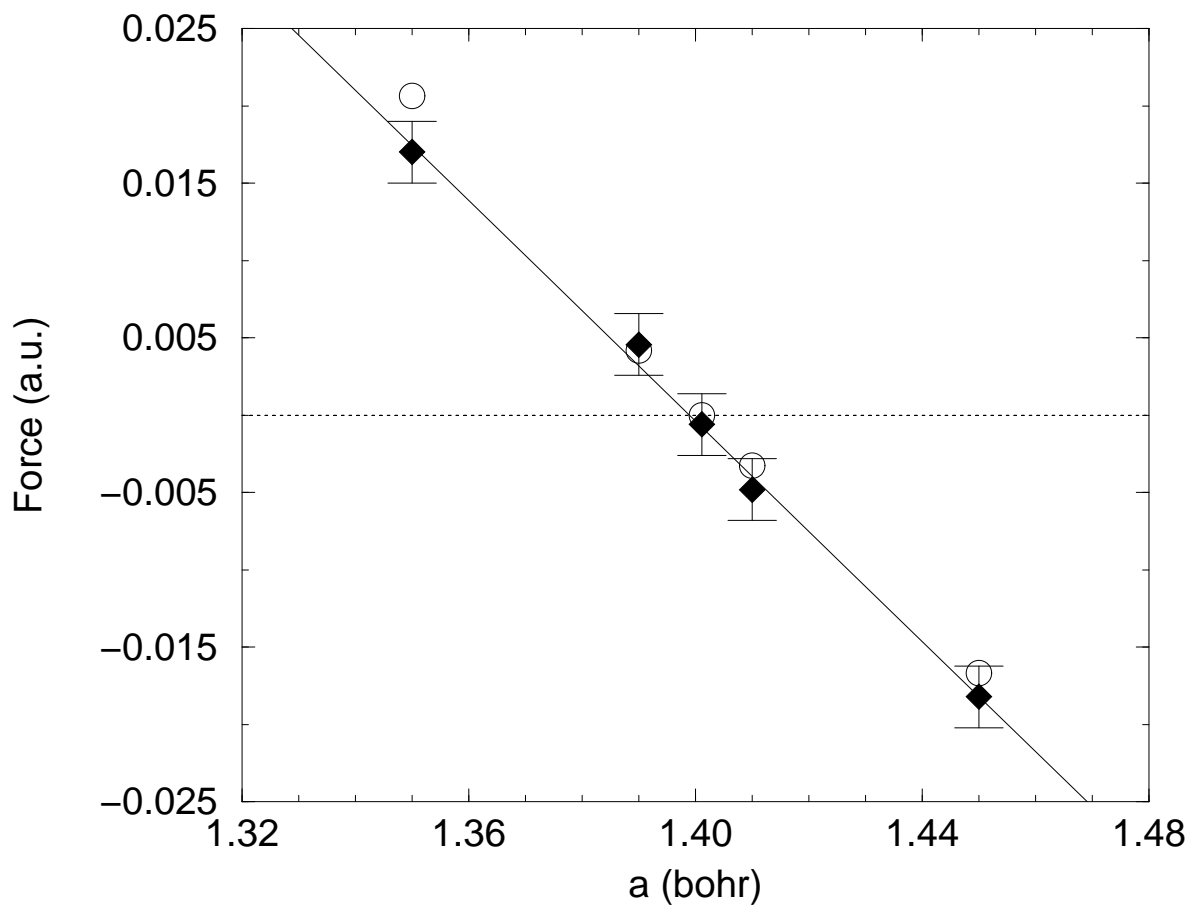}
\end{center}
\caption{The force between two protons in a H$_2$ molecule as a function
of inter-proton distance $a$. The force is plotted as positive when
it is repulsive. Diamonds with error bar are PIMC
results at $\beta=38.4$ with 196 time slices.
Open circles are ground state calculation by
Kolos, et al.\protect\cite{kolos}. 
The solid line is a linear fit to the PIMC
result. The H$_2$ bond length is estimated to be
$1.399\pm .004$.}
\label{H2force}
\end{figure}

Fig. (\ref{H2force}) shows the force between the two protons as a function of
inter-proton distance. Throughout the paper, 
we will use the convention that the force between
the two protons is positive if repulsive.
Very good agreement is obtained with the essentially exact
ground state calculations.
We fit the forces to a straight
line and determined the bond length (which corresponds to $\vec F=0$) of
$1.399\pm 0.004$ bohrs, while the ``exact'' 
ground state value is $1.401$\cite{kolos}. 
The slope of
the line gives the force constant of $0.356\pm 0.028$, in agreement
with the ground state calculation of $0.37$\cite{sun}.

\begin{figure}
\begin{center}
\leavevmode
\epsfbox{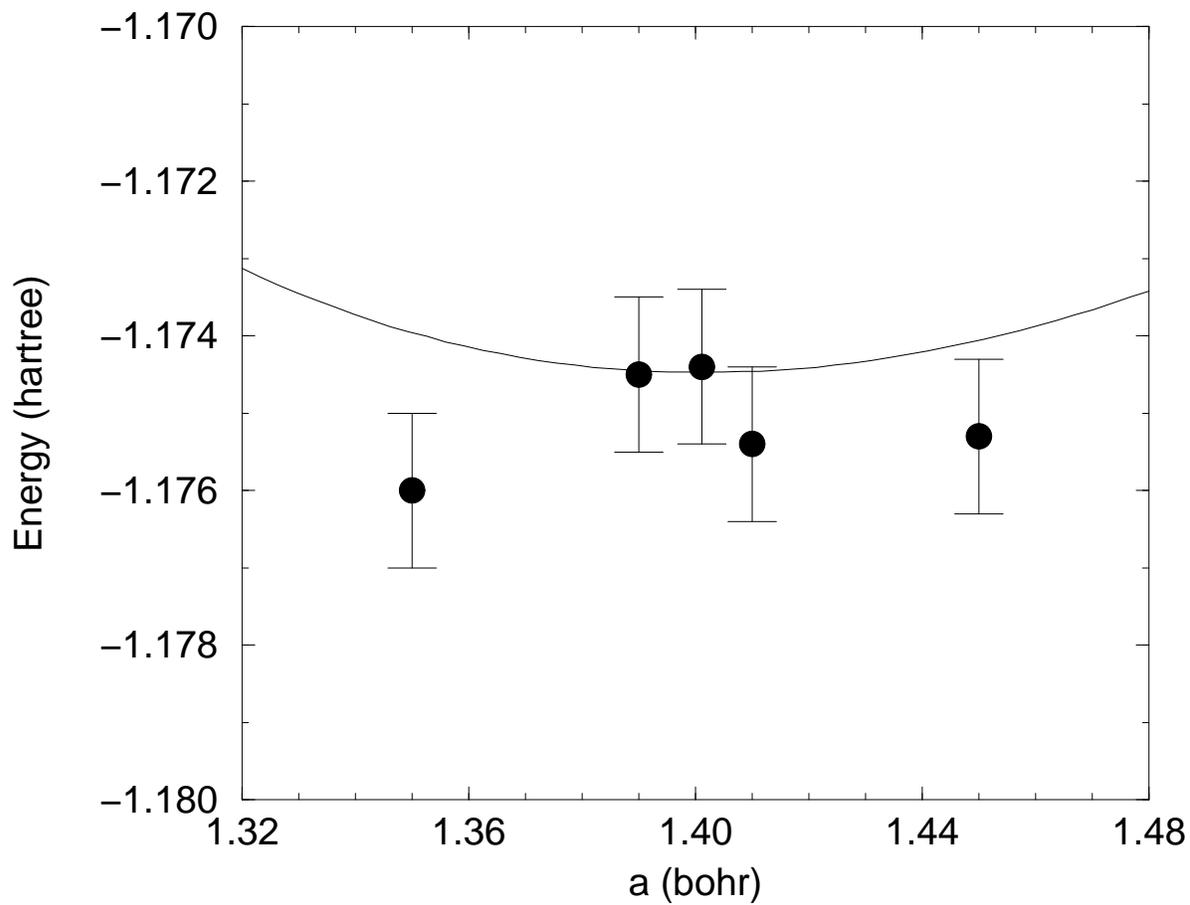}
\end{center}
\caption{The Born-Oppenheimer energy of a H$_2$ molecule by PIMC
simulations are obtained from the same PIMC runs as in 
Fig. (\protect\ref{H2force}).
Time step error was corrected for by extrapolating to the $\tau=0$ limit.
The solid line is ground state calculation by Kolos, et.
al.\protect\cite{kolos}.
}
\label{H2energy}
\end{figure}

In Fig. (\ref{H2energy}) is plotted the total energy around
the equilibrium position from the same PIMC run. 
Note here, in order to compare with ground state energy calculations,
we corrected the time step error by extrapolating to the $\tau=0$ limit.
This is a constant shift of the energy and is unimportant in calculating the
force. Clearly, the total
energy is very flat, and the dependence of the energy on distance is 
completely dominated by the noise.
The reason is two-fold. Firstly, the PIMC force estimator
Eq. (\ref{4}) has a lower variance than the finite difference estimator 
Eq. (\ref{1}). Secondly, the energy is at its
minimum, while the force is not and hence changes more rapidly.
The extremum of the energy (minimum, maximum or saddle points) are
physically important, thus it is important to be able to calculate the
force at those points.

\begin{figure}
\begin{center}
\leavevmode
\epsfbox{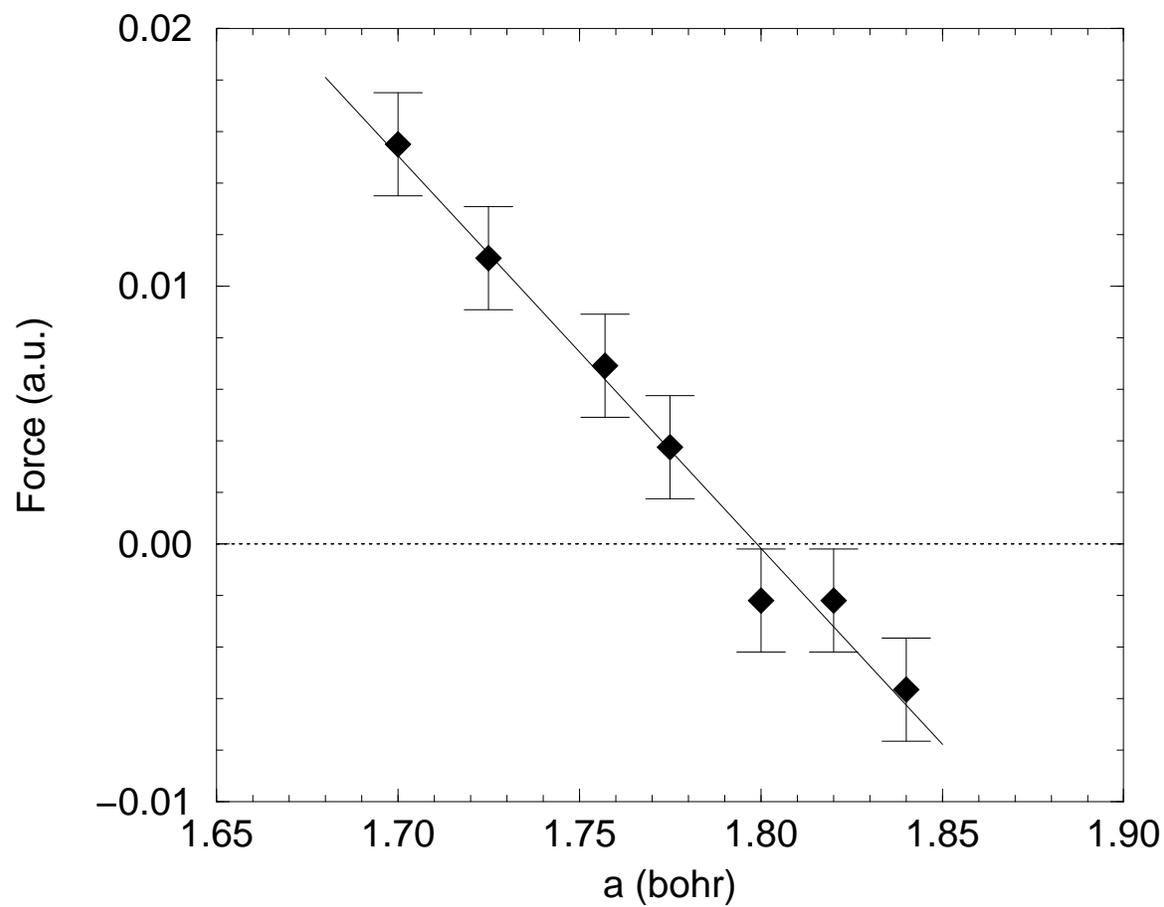}
\end{center}
\caption{The force on the outer protons in a H-H-H collinear
configuration of H$_3$ molecule as a function of the 
nearest neighbor proton-proton distance 
$a$. Diamonds with error bar are PIMC simulations at $\beta=38.4$ with 196 time
slices. The solid line is a fit to the PIMC result.}
\label{H3force}
\end{figure}

The system of H$_3$ does not form a stable molecule.
The interest in this system comes from the need to
determine the barrier of the chemical reaction H+H$_2 \longrightarrow$
H$_2$+H. This is one of the simplest chemical reactions.
It has been found that the barrier occurs at a collinear
configuration of H-H-H with an equal nearest-neighbor separation
$a_1=a_2=a=1.757$\cite{anderson}. 
For a three electron system like H$_3$, fermi
statistics for spin {1/2} particles are needed. 
Anderson has discussed a cancellation scheme for this
system\cite{anderson} in the ground state which has no fixed-node error.
We used the RPIMC with non-interacting
electron nodes to study the collinear H-H-H configuration with 
equal bond distances $a$. 
We used a time step of $\tau=0.2$ with $196$ time slices.
Fig. (\ref{H3force}) 
shows the force on the outer proton with
respect to $a$. The position of the barrier corresponds to the point
where the force is zero (it is a saddle point of the potential energy 
on the $a_1-a_2$ plane), and
we determine it to be at $a=1.799\pm 0.016$ at a temperature of $0.026$
hartrees. Despite the error due to our assumed nodal restriction which does
not have a nodal force and the fact that we are simulating
at a non-zero temperature,
this result compares well with
ground state calculations which give $a=1.757$.

\section{H$_2$ in an Electron Gas}

Hydrogen impurities in metals have attracted much interest, both
experimentally and theoretically\cite{noskov,noskov89,chris,perrot,%
perrot93}. 
In the simplest model of this system, 
one replaces the surrounding metal
ions by a uniform positive charge background and keeps only the valence
electrons.
The lattice structure of the ions and effects of core electrons are
neglected. One studies the energy or equivalently the forces between two
protons as a function of their separation.
The electron screening of the protons is important particularly at low
electron densities. 
Such a model is a good approximation for simple metals 
like Na, Mg and Al
where the conduction electrons are in delocalized plane wave states.

The presence of protons provides particular difficulty for theoretical
studies. Due to the lack of core electrons in a hydrogen atom, 
the electrons see the ``bare'' proton charge. Consequently,
linear response theory is inaccurate\cite{popovic}. 
N\o skov and his coworkers studied
the H-H interaction in an electron gas, particularly, for short
H-H distances ($a<2$ bohrs), with a variety of methods based on the
density functional formalism and the local density approximation. 
First\cite{noskov}, N\o skov solved the equivalent
Dyson equation of the Green function $G$ of the system by projecting out
the difference $\Delta G$ due to the presence of 
H impurities onto a finite local basis set.
Later, he solved the same problem with an Effective
Medium Theory\cite{noskov89} and a self-consistent LDA calculation
\cite{chris}. Perrot\cite{perrot,perrot93} studied
the long range part of the interaction and found indeed, the H-H
potential oscillates at large inter proton distances, $a$, 
due to the Friedel oscillations of the
electron screening. Perrot\cite{perrot93} 
also proposed another method to calculate the
short-range H-H interactions. He considered the molecular binding of
H$_2$ at short $a$ to be the dominant effect, and then determined the
correction due to the electronic density. 

All of the above approaches
give generally consistent descriptions of the proton-proton interaction, 
however, their results are
quantitatively different.
The calculation of the same problem with the
PIMC could be a good test of the validity of the different theoretical
approaches used above. It is also a good test of the algorithm proposed
here 
when applied to an extended system. There are no other methods that can
easily carry out such a calculation.

\begin{figure}
\begin{center}
\leavevmode
\epsfbox{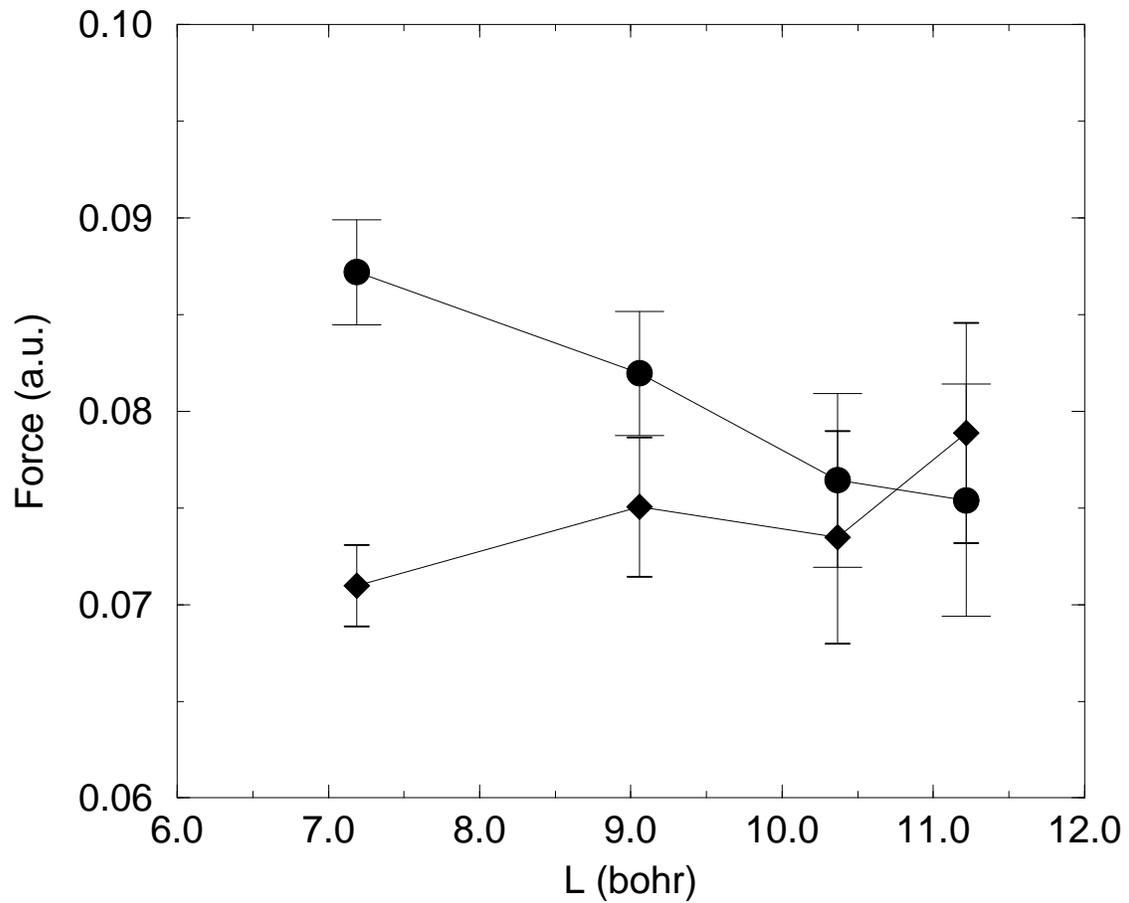}
\end{center}
\caption{The dependence of the proton-proton interaction in an electron
gas on the size of the simulation cell. $\beta=16.0$ and the time step
$\tau=0.4$. The proton-proton separation is $1.5$ bohrs.
The electronic density corresponds to $r_s=2.07$.
The number of electrons in the lower curve are determined by 
Eq. (\protect\ref{5}) and
have slightly lower density. The upper curve has two more electrons in
the cell.}
\label{size}
\end{figure}

Jellium is characterized by the dimensionless Wigner sphere
radius $r_s$, defined by $r_s=(3/4\pi n)^{1/3}$ with $n$ the
electronic density. We carried out PIMC simulations of H$_2$ in an
electron gas of densities representing Na ($r_s=3.93$) and Al ($r_s
=2.07$) and computed the force between the two protons. The constant
background of ions gives no contribution to this force. Ewald sums are
used to compute the long range contributions from periodic images of
proton and electron charges.
Our supercell
consisted of two protons with a fixed separation of $a$ and
$N_e$ electrons in a cube of size $L$ with 
\begin{equation}\label{5}
N_e=\frac{3}{4\pi}\left (\frac{L}{r_s}\right )^3.
\end{equation}
Since some
electrons could be bound to the protons and occupy a relatively smaller
space, the actual surrounding electronic density is lower than
$n$. However, this finite-size correction decreases rapidly as $L$ increases
as shown in Fig. (\ref{size}) and discussed below. In most of the
following,
we used a temperature of $T=1/16$ hartrees with two time steps of:
$\tau=0.4$ and $\tau=0.8$. The $\tau=0$ limit is extrapolated by fitting
the result to $F(\tau)=F(0)+\alpha\tau^2$. A typical calculation with
38 electrons and $\tau=0.4$ takes about 100 CPU hours on a SGI/CRAY
Origin2000 computer.

Fig. (\ref{size}) shows the convergence of the result as the cell size increases.
For the lower curve (diamond), the number of electrons is determined by
Eq. (\ref{5}).
For the upper curve (circle), two additional electrons are added to the
system so that the surrounding electron density is higher than $n$. The
two results should bound the result for density $n$ from above and
below.
As $L$ increases, the
difference between the two curves quickly reduces and at $L\geq 10$, no
noticeable size dependence is present within error bars.

\begin{figure}
\begin{center}
\leavevmode
\epsfbox{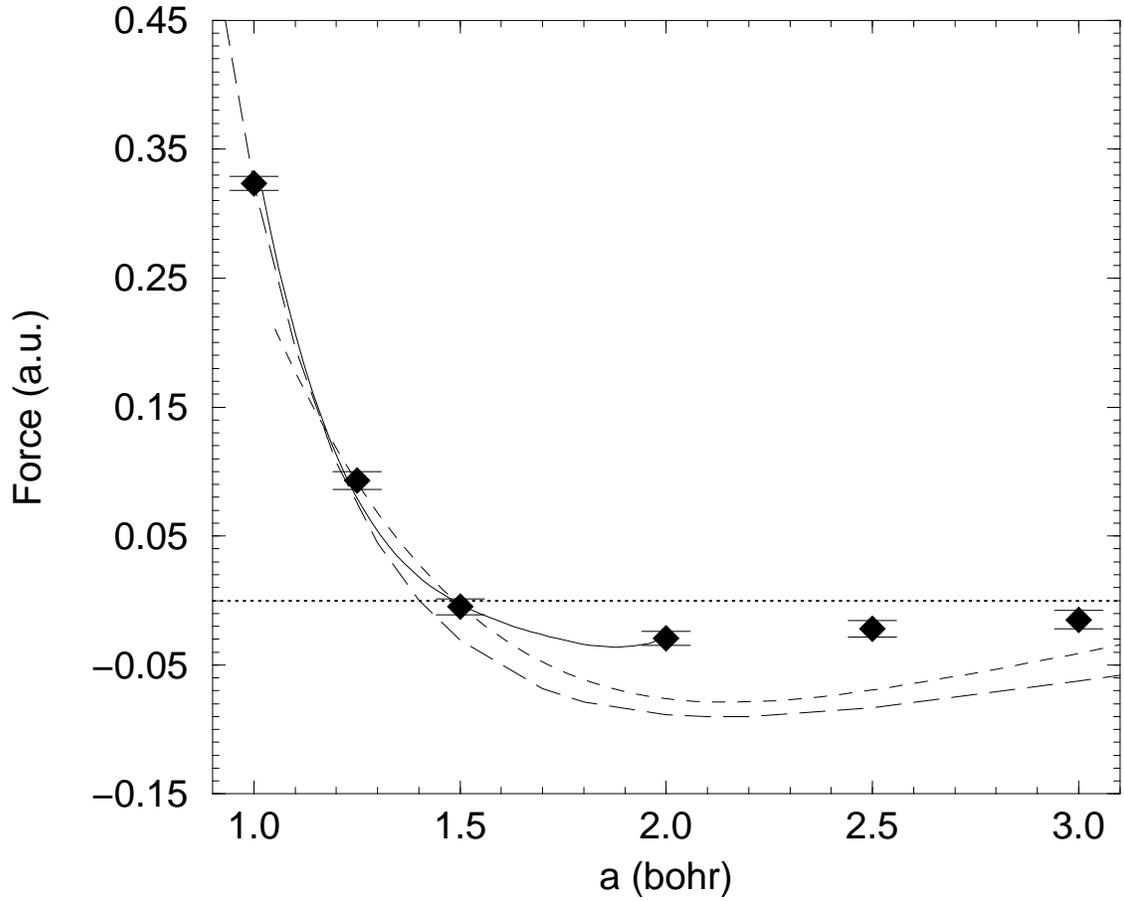}
\end{center}
\caption{The proton-proton interaction in an electron gas with
$r_s=3.93$. The diamonds with error bars are PIMC results at $\beta=16$.
Two time step values $\tau=0.4$ and $0.8$ were used and the results
extrapolated to $\tau=0$. 38 electrons were in the simulation cell.
The solid curve is LDA calculations by N\o skov\protect\cite{noskov}
and the dashed curve is by Perrot\protect\cite{perrot93}. 
They were obtained by a fifth
degree polynomial fit to the binding energy data and then taking an
analytic derivative. For comparison, the proton-proton force in a H$_2$
molecule\protect\cite{kolos} is also plotted as the 
long dashed curve.}
\label{3.93}
\end{figure}

\begin{figure}
\begin{center}
\leavevmode
\epsfbox{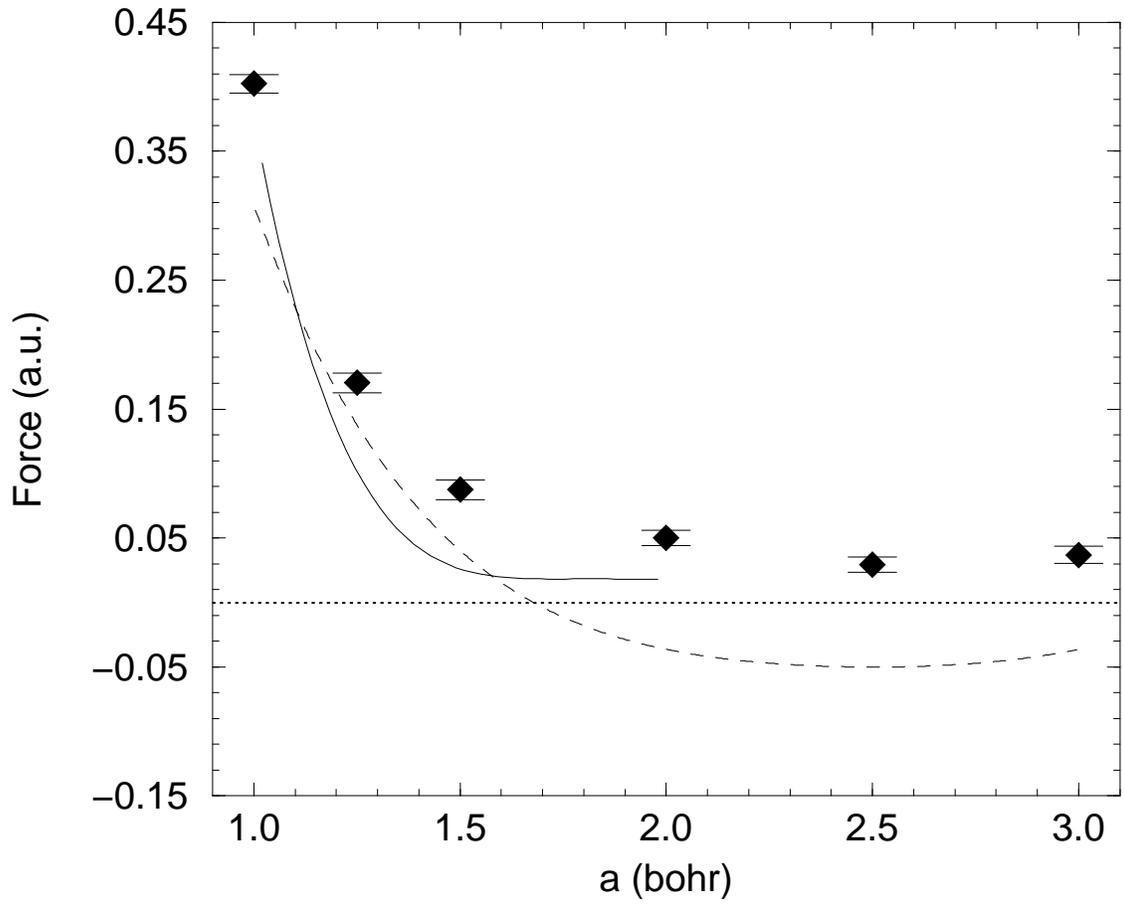}
\end{center}
\caption{The proton-proton interaction in an electron gas with
$r_s=2.07$. 
The solid curve is LDA calculations by N\o skov\protect\cite{noskov}
and the dashed curve is by Perrot\protect\cite{perrot93}. 
Other details are as in Fig. (\protect\ref{3.93}).}
\label{2.07}
\end{figure}

We performed calculations with $N_e=38$ ($L=21 $ for $r_s=3.93$ 
and $L= 11$
for $r_s=2.07$) which is shown by tests to be large enough that the finite size
effect is small with respect to the error bars. In Fig. (\ref{3.93}) and
Fig. (\ref{2.07}), we plot the
result against previous calculations of N\o skov\cite{noskov} and Perrot
\cite{perrot93}. 
The presence of an electron gas greatly reduces the H-H binding
with respect to a free H$_2$ molecule. At $r_s=3.93$ (Fig. \ref{3.93}),
where the electron
density is low, the short range force is still dominated by the bonding
electrons and it is almost the same as that of a free H$_2$. As $a$
increases, the strong attractive force in a H$_2$ molecule becomes
substantially weaker. The equilibrium H-H position is $a\approx
1.5$, larger than the H$_2$ bond length.
One explanation is that the
anti-bonding state of the H-H system is also partially filled when it is
placed in an electron gas. 
N\o skov's LDA calculation agrees quite well with the PIMC calculation.
The PIMC extends easily to large proton separations 
while N\o skov's
method has an increasing numerical error at large $a$ due to his choice of the
one-centered basis functions
and is limited to $a<2$.
For a higher density, $r_s=2.07$ (Fig. \ref{2.07}), 
the H-H interaction is
completely repulsive. N\o skov's calculation, though giving the same
qualitative description, is less repulsive than our PIMC result. This may be
caused by the lack of the electron-electron correlation in the LDA.
Perrot's calculation, which always predicts
the existence of an equilibrium H-H distance even for densities as
high as $r_s=2.07$, does not agree with either the LDA or PIMC results. 

We also studied the temperature effects on the electronic screening for
$a=2.0$. 
At low temperatures, electrons are bound to the protons and
provide an attractive force which overcomes the Coulomb repulsion between
the two protons. As the temperature increases, the electrons become more
energetic and it is more and more difficult to confine them to the
vicinity of protons. At sufficient high temperature, electrons are almost
free, we recover
the pure proton-proton Coulomb force($1/a^2$).

\begin{figure}
\begin{center}
\leavevmode
\epsfbox{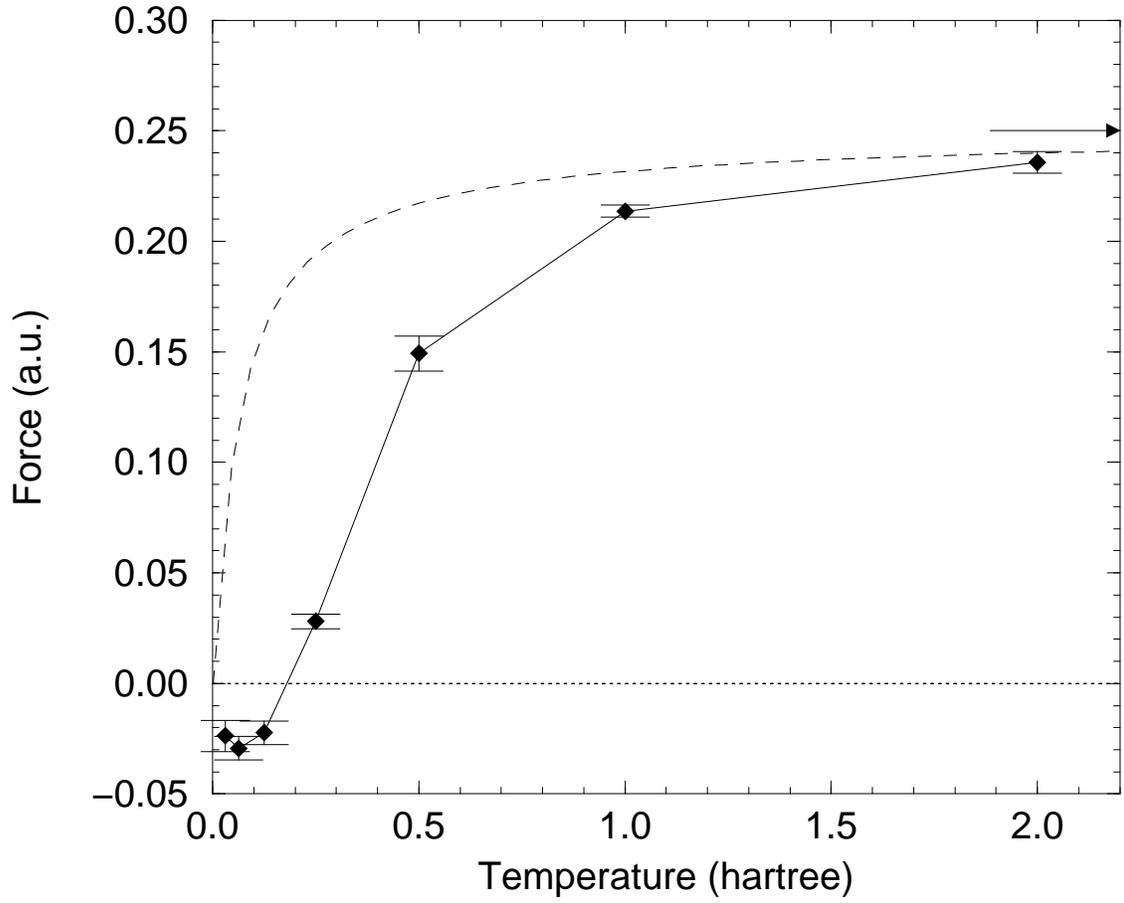}
\end{center}
\caption{Temperature dependence of the proton-proton interaction in an
electron gas with $r_s=3.93$ at a proton separation of $a=2.0$. Diamonds
with error bars and the solid curve are PIMC results. 
The dashed curve is from the
Debye model. The arrow at $F=0.25$ a.u. indicates the high
temperature limit.}
\label{temp}
\end{figure}

In the non-degenerate limit ($n\lambda^3\ll 1$ where
$\lambda=\hbar/\sqrt{m_ek_BT}$), the electronic screening can be
understood by the Debye theory\cite{ebeling}, 
where the screened proton potential is:
\begin{equation}
v_s(r)=\frac{e}{r}\exp(-r/r_0)
\end{equation}
and $r_0=(4\pi e^2\beta n)^{-1/2}$ is the Debye length.
As Fig. (\ref{temp}) shows, at very high temperature ($T\geq 2$ hartrees), 
the result of Debye screening agrees with that of
the PIMC calculations. As temperature decreases, the Debye picture fails
both because of the strong degeneracy of electrons and most importantly,
the formation of electronic bound states around the protons.

\section{Outlook}

In this paper, we presented a method to calculate the electronic forces
with PIMC simulations. There is no trial function involved in PIMC which
makes it much easier to simulate a variety of physical systems at different
geometries. The force estimator is local and thus it is
both easy to calculate and scales well with the system size.
Temperature is included naturally in the PIMC. This could be a
disadvantage, for example, for the study of bond breaking at very low
temperature. But it also allows
one to study the finite temperature behavior of the system, as well as
comparing with experiments directly. Finally and most importantly, 
electronic correlations are
included so that one could study systems where the correlation is
strong.

We demonstrated the effectiveness of this method by applying it both to
simple molecules and an extended electron gas. The approximation of the
fermion nodal
surface is the only uncontrolled approximation in this approach. This
approximation could be significant when bound states form. 
The variational path integral technique\cite{cep95} would allow one to
perform truly ground state force calculations using this method, at the
expense of reintroducing a trial function.
A natural extension of this work is to study the effects of nodes that also
depend on nuclear coordinates. For example, one could use 
the nodal surface resulting
from a one-particle self-consistent Hartree-Fock or LDA calculations
\cite{cep96}. One could also put in a ``backflow'' effect by transforming
paths onto ``quasi-particle'' coordinates. This approach has been found
to be very successful for liquid $^3$He\cite{panoff} and $2D-3D$ electron 
gas\cite{kwon}.
It would be interesting
to develop a Monte Carlo method similar to the Car-Parrinello
approach with atoms moving on the Born-Oppenheimer potential energy
determined by QMC calculations once the force calculation can be done
efficiently enough.

\acknowledgments

This work was supported by NSF grant No. DMR-94-224-96, ONR grant No.
N00014-92-J-1320 and the Department of Physics at the University of
Illinois at Urbana-Champaign. 
Calculations were performed at the National Center for
Supercomputing Applications.

\end{document}